\documentclass[12pt,prd,superscriptaddress,nofootinbib]{revtex4}
\usepackage[english]{babel}
\usepackage{graphicx}
\usepackage{mathtext}
\usepackage{indentfirst}
\usepackage{epsfig,amsmath,amsfonts}

\def\ee{\end{equation}}
\def\ba{\begin{eqnarray}}
\def\ea{\end{eqnarray}}

\def\bq{\begin{quote}}
\def\eq{\end{quote}}

 at 10truept

\newcommand{\beq}{\begin{equation}}
\newcommand{\eeq}{\end{equation}}
\newcommand{\beqa}{\begin{eqnarray}}
\newcommand{\eeqa}{\end{eqnarray}}
\newcommand{\bea}{\begin{eqnarray}}
\newcommand{\eea}{\end{eqnarray}}

\def\ltap{\ \raise.3ex\hbox{$<$\kern-.75em\lower1ex\hbox{$\sim$}}\ }
\def\gtap{\ \raise.3ex\hbox{$>$\kern-.75em\lower1ex\hbox{$\sim$}}\ }
\def\gl{\ \raise.5ex\hbox{$>$}\kern-.8em\lower.5ex\hbox{$<$}\ }
\def\roughly#1{\raise.3ex\hbox{$#1$\kern-.75em\lower1ex\hbox{$\sim$}}}

\newcommand\bqa {\begin{eqnarray}}
\newcommand\eqa {\end{eqnarray}}
\newcommand{\bec}{\begin{cases}}
\newcommand{\eec}{\end{cases}}
\newcommand{\bei}{\begin{itemize}}
\newcommand{\eei}{\end{itemize}}
\newcommand{\bee}{\begin{enumerate}}
\newcommand{\eee}{\end{enumerate}}

\newcommand\pr {\partial}
\newcommand{\fr}{\frac}

\newcommand\nn {\nonumber}

\newcommand{\bear}{\begin{array}}
\newcommand{\enar}{\end{array}}

\begin{document}

\def\I{{\rm i}}

\def\h{\hbar}

\def\t{\theta}
\def\T{\Theta}
\def\w{\omega}
\def\ov{\overline}
\def\a{\alpha}
\def\b{\beta}
\def\g{\gamma}
\def\s{\sigma}
\def\l{\lambda}
\def\wt{\widetilde}
\def\t{\tilde}

\hfill ITEP--TH--43/12

%\hfill AEI--2012--011

\vspace{5mm}

\title{\Large Physical meaning and consequences of the loop IR divergences in global dS}

\author{\bf E.\ T.\ Akhmedov}
%\email{akhmedov@itep.ru}
\affiliation{B.\ Cheremushkinskaya, 25, Institute for Theoretical and Experimental Physics, 117218, Moscow, Russia}
\affiliation{Institutskii per, 9, Moscow Institute of Physics and Technology, 141700, Dolgoprudny, Russia}
\affiliation{Vavilova, 7, Mathematical Faculty of the National Research University Higher School of Economics, 117312, Moscow, Russia}

\maketitle

\begin{center}{\bf Abstract}\end{center}
Following arXiv:1012.2107 we show that in global de Sitter space its isometry is broken by the loop IR divergences for any invariant vacuum state of the massive scalars. We derive kinetic equation in global de Sitter space. It follows from the Dyson-Schwinger equation of the Schwinger-Keldysh diagrammatic technique in IR limit and allows to understand the physical meaning and consequences of the loop IR divergences. In many respects the isometry breaking in global dS is similar to the one in the contracting Poincare patch of de Sitter space. Hence, as a warm up exercise we study the kinetic equation and properties of its solutions in the expanding and contracting Poincare patches of de Sitter space. Quite unexpectedly we find that under some initial conditions there is an explosive production of massive particles in the expanding Poincare patch.

\vspace{5mm}

\section{Introduction}

De Sitter's (dS) space isometry is crucial for its stability. Obviously if it is not broken, then the cosmological constant can not be secularly screened. There are states of the massive scalar field theories in which they respect dS isometry at tree level \cite{Mottola:1984ar},\cite{Allen:1985ux}. They are referred to as $\alpha$--vacua. But in the loops the situation becomes more complicated. One loop calculation in the expanding Poincare patch (EPP) of dS space shows that the isometry is broken for all $\alpha$--vacua except Bunch-Davies (BD) one \cite{Polyakovtalk} (see also the Appendix of \cite{Akhmedov:2012pa}).

However, we find it as rather unphysical to address the question of the stability of the system under such conditions when all its symmetries are respected exactly. We propose to consider excitations above the highly symmetric state and to trace where do they evolve in the future infinity.
For the finite particle density on top of any $\alpha$-vacuum (including BD one) dS isometry is broken even at tree-level \cite{Akhmedov:2012pa},\cite{Polyakovtalk}. The obvious question to ask at this point is whether this density decays or rapidly increases with time?
Rephrasing this, we would like to understand if the system has the dS invariant ground state.

The standard answer to this question is that the density will decay due to the rapid expansion of the space. But the situation happens to be more complicated and actually quite counterintuitive. We show in this paper that the proper quantity to study is the number density per comoving volume: It is crucially important and may be counterintuitive that the particle creation and decay rates are defined, at the non-linear level, via the density per comoving volume rather than per physical one. (Unlike the physical volume, which inflates with time, comoving volume remains constant.) Note that the number density per comoving volume of the spatially homogeneous dust remains constant throughout the history of dS space independently of whether the space is expanding or contracting. But such a density will grow unboundedly if one will turn on only the process of particle production by the gravitational field. That is, in particular, the source of the loop IR divergences and large IR contributions, which are under study in this paper. Of cause for the issue of the backreaction it is important to know how fast is such a growth in comparison with the expansion of the space, especially when the loss processes are also taken into account (on top of the production ones)\footnote{Note that turning on only single process of the particle kinetics, rather than all of them, may cause an instability, i.e. may cause a divergence in the collision integral (see \cite{Polyakovtalk} for the detailed discussion). However, as we show, in many cases these instabilities do disappear when one takes into account all kinetic processes together. I.e. collision integral becomes well defined.  But then, in turn, new instabilities, of different kinds, which cause an explosion of the solution of the kinetic equation (but not of the collision integral), may appear under some conditions.}. The main goal of this paper is to address this problem.

In global dS the situation is more intricate. There is one-loop IR divergence even for massive scalars \cite{Krotov:2010ma}. It reveals itself via impossibility to take the moment of turning on selfinteractions, $t_0$, to past infinity. We are going to see that $t_0$ plays the role of IR cutoff. Thus, unlike the situation with the BD initial state in EPP, in global dS isometry is broken in the loops for all possible vacua (including Euclidian one). Furthermore, the presence of such IR divergences means that the growth of particle density wins against the expansion of the space--time.

However, it remains unclear whether after the summation of all loop leading IR contributions one can take $t_0 \to - \infty$ or cannot. We show that this problem is related to the one of the presence of a stationary state in dS QFT. If such a state does exist, then $t_0$ can be taken to past infinity and the dS invariance is restored. Otherwise the dS isometry is not restored at higher loops and the backreaction due to the particle production cannot be neglected. It was explained in \cite{Akhmedov:2011pj,Akhmedov:2012pa} that to approach such a problem one has to solve the kinetic equation, which is the tool for the summation of the leading IR contributions in all loops. Below we derive this equation in global dS.

Although in EPP there are no IR divergences, one has there large IR contributions. As the result, one can also derive the kinetic equation in EPP for the finite initial particle density on top of the BD vacuum. In \cite{Akhmedov:2011pj,Akhmedov:2012pa} the problem of the existence of the stationary state in EPP was addressed with the use of the corresponding kinetic equation. The conclusion was that it can exist under the appropriate initial conditions.

In the contracting Poincare patch (CPP) the situation happens to be similar to the global dS. There, unlike EPP, we also encounter loop IR divergences. In the main body of this paper we explain the reason of such a difference between EPP and CPP. We derive the kinetic equation in CPP. Such equations in EPP and CPP can be simply mapped to each other, because we are dealing with the spatially homogeneous states and study the density per comoving volume. As the result there is a quite simple map between the solutions of the kinetic equations in EPP and CPP. In particular, we find that there is the stationary solution even in the CPP, which is related to the above mentioned stationary solution in EPP. Its presence means that under the appropriate initial conditions and in the circumstances of the perfect spatial homogeneity, dS isometry can be restored, after the summation of all leading loop contributions, even in CPP.

But one of the main messages of this paper is that there is also a solution of the kinetic equation (both in CPP and EPP) which describes explosion, within a finite proper time, of the particle density. The presence of such a solution means that under some initial conditions backreaction from the particle production can become enormous even in EPP. The last type of the situation should lead to the deformation of the dS metric and probably bring the system to the FRW space-time with secularly decreasing cosmological constant.

\subsection{Subtleties}

To proceed let us point out some subtleties with field theory in curved space-times, in general, and in dS space, in particular. For illustrative reasons consider 4D massless conformaly coupled scalar field with the quartic selfinteraction, $\left[\Box + R/6\right] \, \phi = \lambda \, \phi^3$, on global dS background. We are interested in the following classical scattering problem. Selfinteraction is adiabatically turned on somewere in the far past and then eventually switched off in the far future. At past infinity we excite a single free wave and would like to ask the following question: will the solution of the equation under consideration (at the linear order in $\lambda$) contain three free waves in future infinity or will not?

In the metric with noncompact spatial slices $ds^2 = \frac{1}{\eta^2}\left[d\eta^2 - d\vec{x}^2\right]$, $\eta \in (-\infty, +\infty)$ (see fig. 1), the problem can be mapped, via conformal transformation, to the one in Minkowski space. The intensity of the process in question, at the leading $\lambda$ order, will be proportional to the $\delta$-function ensuring energy conservation \cite{Akhmedov:2008pu}. The latter will put an obstruction to the process under consideration.

If, however, one will address the same problem in the metric $ds^2 = dt^2 - \cosh^2(t)d\Omega^2 = \frac{1}{\cos^2(T)}\left[dT^2 - d\Omega^2\right] $ (see fig. 2), then it can be converted to the one in Einstein static universe with the {\it compact} range of time, $T \in [-\frac{\pi}{2}, +\frac{\pi}{2}]$. As a result instead of the aforementioned $\delta$-function one will encounter $\frac{\sin[\pi \Delta E]}{\Delta E}$, where $\Delta E$ is the energy deficit in the process in question. Hence, the process will be allowed.

The resolution of the apparent disagreement with the general covariance is that the result of this scattering process strongly depends on the initial/boundary conditions. Approaching the same problem with the use of different ways of spatial slicing of dS space we implicitly assumed different initial conditions. In fact, while in the case of noncompact spatial sections one has such harmonics as $e^{-i \, \omega \, \eta}$, for the situation with the compact sections the harmonics are $e^{-i\, \epsilon \, T}$. The map between these harmonics under the coordinate and conformal transformations should be of the main concern for those who are studying the scattering problem under consideration.

The observations of the above paragraphs are relevant for us because below we are going to study IR contributions, which of cause are sensitive to the conditions at the boundaries of various patches of the entire space-time.
Also they give one a hint for the explanation of the disagreement between the conclusions of \cite{MarolfMorrison} and those of the present paper. The spatial slicing of fig. 1 was used in \cite{MarolfMorrison}. Although in the latter paper the entire dS space was considered, the BD state was imposed at the boundary separating contracting CPP and EPP. We do not quite understand the physical meaning of this approach, but we can see that in such a way it is possible to respect dS isometry at each step of the construction of the correlation functions. In fact, as we have already pointed out above, to respect dS isometry in all loops one has to impose BD state exactly at the boundary separating EPP from CPP.

At the same time, in the kinetic theory approach one can use both spatial slicings of fig. 1 and fig. 2. In this setting there is nothing behind the initial Cauchy surface and the dS isometry is broken, at least in the loops, unless we start with the exact BD state exactly at the boundary of EPP.

\begin{figure}[t]
\begin{center} \includegraphics[scale=0.3]{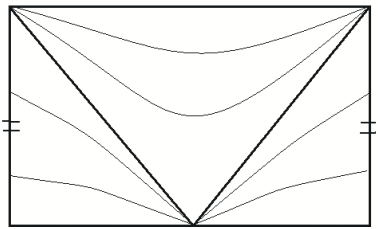}
\caption{Penrose diagram of dS space. Left and right sides of the depicted rectangle are glued to eachother. Here we show Cauchy surfaces (e.g. constant $\eta$ slices) for the global dS metric of the form $ds^2 = \frac{1}{\eta^2}\left[d\eta^2 - d\vec{x}^2\right]$, where $\eta \in (-\infty, +\infty)$.}
\end{center}
\end{figure}

\begin{figure}[t]
\begin{center} \includegraphics[scale=0.3]{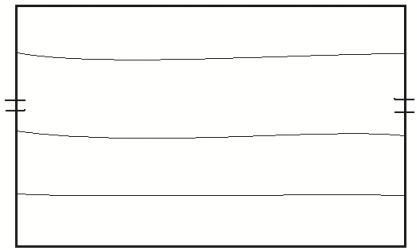}
\caption{Here we show Cauchy surfaces (e.g. constant $t$ or $T$ slices) for the global dS metric of the form $ds^2 = dt^2 - \cosh^2(t)d\Omega^2 = \frac{1}{\cos^2(T)}\left[dT^2 - d\Omega^2\right] $, where $t \in (-\infty, +\infty)$ and $T \in [-\frac{\pi}{2}, +\frac{\pi}{2}]$.}
\end{center}
\end{figure}

\subsection{The setup of the problem}

Our eventual goal is to see whether the back--reaction from the particle production in dS space is negligible or not. But in this paper we neglect back--reaction and consider self--interacting massive scalar field theory on the fixed dS background. We restrict our attention to the behavior of the solutions of the kinetic equation in the IR limit and try to see if the assumption of the negligible backreaction is self consistent. Note that in EPP (CPP) the IR limit of the physical momentum corresponds to future (past) infinity.

We study $D$-dimensional minimally coupled real scalar field theory:

\bqa\label{FT}
L = \sqrt{|g|}\,\left[\frac{g^{\mu\nu}}{2}\, \pr_\mu \phi \, \pr_\nu\phi + \frac{m^2}{2}\, \phi^2 + \frac{\lambda}{3}\, \phi^3 + \dots\right].
\eqa
Dots here stand for the higher power self--interaction terms, which make the theory stable.
However, below we are going to consider formulas only due to the unstable cubic part of the potential. The reason for that is just to simplify equations. This instability does not affect our conclusions, because similar IR contributions do also appear in massive $\phi^4$ theory \cite{Akhmedov2} or in the other nonconformal theories in dS space \cite{Woodard}, \cite{Dolgov:1994cq}, \cite{Antoniadis:2006wq}, \cite{Xue:2012wi}, \cite{Giddings:2010ui}.

To set the notations we describe here some of the properties of this space and of free scalar harmonics on it. It is the hyperboloid $X_0^2 - X_i^2 = - 1$, $i = 1, \dots, D$ in the $(D+1)$--dimensional Minkowski space $ds^2 = dX_0^2 - dX_i^2$. Throughout this paper we set the curvature of the hyperboloid to one. The dS isometry group is $SO(D,1)$ Lorentz group of the ambient Minkowski space in question. Throughout this paper we assume that $D$ is even. For odd $D$ the situation is a bit more subtle, but still is not much different \cite{Polyakovtalk}.

The global dS metric is induced through the solution of the equation $X_0^2 - X_i^2 = - 1$ via $X_0 = \sinh t$ and $X_i = \omega_i \, \cosh t$, where $\omega_i^2 = 1$ defines the coordinates on the $(D-1)$--dimensional sphere. The induced metric is then $ds^2 = dt^2 - \cosh^2 t \, d\Omega^2$.

The EPP is the halfspace of this hyperboloid. It is defined by the condition $X_0 \geq X_D$. Any other EPP is obtained by a rotation of the latter patch around $X_0$ axis, i.e. there is a family of EEPs, which is parametrized by the angle of the rotation around $X_0$ axis. Such a patch of dS respects only some subgroup of the whole isometry group. Its induced metric is $ds^2 = dt^2 - e^{2t}\, d\vec{x}^2 = \frac{1}{\eta^2} \, \left[d\eta^2 - d\vec{x}^2\right]$, where $\eta = e^{-t}$ is the conformal time, which is flowing in the reverse direction --- from infinity (in the past) to zero (in the future).

The CPP also is a halfspace of the hyperboloid, which is defined e.g. by the condition $X_0 \leq X_D$, i.e. it is complementary to EPP within the global dS. The induced metric on the CPP is $ds^2 = dt^2 - e^{-2t}\, d\vec{x}^2 = \frac{1}{\eta^2} \, \left[d\eta^2 - d\vec{x}^2\right]$, where $\eta = e^{t} \in [0,+\infty)$.

For large enough $D$ the theory (\ref{FT}) becomes non--renormalizable. However, as we explain in the concluding section, if there is a stationary state in this theory, then in the IR limit we do not care about UV divergences and renormalizability of the theory in question. We assume that all couplings in all equations take their physical values, i.e. all UV divergent ($\sim \lambda^2 \, \log \Lambda$) or finite ($\sim \lambda^2$) contributions are absorbed into their renormalization. It is straightforward to do that during the derivation of the kinetic equation from the Dyson--Schwinger one. At this point one should keep in mind that for the propagators which have proper Hadamard behavior the UV divergences in dS space are the same as in flat one. Only BD harmonics provide the proper Hadamard behavior in EPP and CPP. In global dS the same is true only for the Euclidian harmonics.

The scalar field harmonics in global dS satisfy the Klein--Gordon equation:

\bqa\label{no2}
\left[\pr_t^2 + (D-2)\, \tanh(t) \, \pr_t + m^2 - \frac{\Delta_{D-1}(\Omega)}{\cosh^2(t)}\right]\phi = 0
\eqa
where $\Delta_{D-1}(\Omega)$ is the Laplacian on the $(D-1)$--dimensional sphere. Any solution of this equation can be expanded in harmonics as follows $\phi_p(t,\Omega) = g_p(t) \, Y_{p\vec{m}}(\Omega)$, where $Y_{p\vec{m}}(\Omega)$ are $(D-1)$--dimensional spherical functions and $g_p(t)$ can be expressed through the hypergeometric function \cite{Mottola:1984ar}, \cite{Allen:1985ux}.

For large enough $p$ we can somewhat simplify eq.(\ref{no2}) in the limit $t\to\pm \infty$.
In fact, under these conditions and after the change of variables $\eta_{\pm} = e^{\mp t}$ the global dS metric can be approximated by $ds^2_{\pm} \approx \frac{1}{\eta_\pm^2} \, \left[d\eta_\pm^2 - d\vec{x}^2\right]$. Here conformal time $\eta_+$ flows towards zero ($\to + 0$) in future infinity and, thus, covers a part of the EPP; similarly $\eta_-$ starts from zero ($0+ \to$) at past infinity, and covers a part of the CPP. Furthermore, the substitution of $\phi(t\to \pm \infty) \approx \eta_{\pm}^{\frac{D-1}{2}}\, h_{\pm}(p\eta_{\pm}) \, e^{\mp i \, \vec{p}\, \vec{x}}$ into the simplified (by the limit $t\to\pm\infty$) form of (\ref{no2}) reduces it to that of Bessel for $h_{\pm}(p\eta)$. The index of the obtained Bessel equation is $i\mu$, where $\mu = \sqrt{m^2 - \left(\frac{D-1}{2}\right)^2}$.

Throughout this paper we are going to work with the massive fields from the principal series, $m > (D-1)/2$. It can be shown \cite{Akhmedov2}, however, that for the fields from the complementary series, $m < (D-1)/2$, one-loop IR divergences have the similar logarithmic character to that of massive fields. However, for the fields from the complementary series it is not clear how to write the kinetic equation of the conventional form, because for them the necessary separation of scales between the time dependence of the harmonics and of the particle density is not obvious.

\section{Contracting Poincare patch}

The reason why we start our discussion with CPP is that the physics of IR divergences there is relevant for the field dynamics in global dS.

To calculate loops in the non--stationary situation one has to use the in-in or Schwinger-Keldysh diagrammatic technique. Feynman rules in EPP can be found in \cite{vanderMeulen:2007ah}. In CPP the rules are very similar.

Each particle in the non--stationary situation is described by the matrix of propagators. The constituents of this matrix are the retarded and advanced propagators $D^{R,A}$, which carry information about the particle spectrum of the theory in question, and the Keldysh propagator $D^K$, which carries the statistical information, i.e. shows which levels are populated.

Due to spatial homogeneity of CPP together with the initial state, we find it convenient to perform the Fourier transform of all quantities along the spatial direction $\vec{x}$: $D^{K,R,A}_p(\eta_1, \eta_2) \equiv \int d^{D-1}\vec{x} \, e^{i\, \vec{p} \, \vec{x}} D^{K,R,A}(\eta_1, \vec{x}; \eta_2, 0)$.
Then at tree level the Keldysh propagator is $D_{0p}^{K}(\eta_1,\eta_2) = (\eta_1\eta_2)^{\frac{D-1}{2}}\,{\rm Re}\left[h(p\eta_1)h^*(p\eta_2)\right]$, retarded and advanced propagators are $D_{0p}^{R}(\eta_1,\eta_2) = \theta\left(\eta_1-\eta_2\right) \,2\,(\eta_1\eta_2)^{\frac{D-1}{2}}\,{\rm Im}\left[h(p\eta_1)h^*(p\eta_2)\right]$ and $D_{0p}^{A}(\eta_1,\eta_2) = - \theta\left(\eta_2-\eta_1\right) \, 2\, (\eta_1\eta_2)^{\frac{D-1}{2}}\,{\rm Im}\left[h(p\eta_1)h^*(p\eta_2)\right]$; $h(p\eta)$ is a solution of the Bessel equation, which we do not specify so far. The choice of the solution defines which harmonic basis we are using to quantize the theory. In other words, it specifies what we have chosen as the background state in our problem\footnote{The choice of any other $\alpha$--vacuum except Bunch--Davies one at past infinity of EPP would affect the UV behavior in the theory. But we are interested in study of the IR properties of the $\alpha$--vacua. At this point one should keep in mind that Bogolubov rotation in the low energy effective BCS theory does not affect the UV behavior of QED.}.

In the limit $\eta = \sqrt{\eta_1\eta_2} \to \infty$, $\eta_1/\eta_2 = const$ and $\eta_0 \to 0$ the leading one-loop IR contribution to $D_p^K(\eta_1,\eta_2)$ is contained within the following expression

\bqa\label{CPP}
D_{1p}^{K}(\eta_1,\eta_2) = (\eta_1\,\eta_2)^{\frac{D-1}{2}}\, \left[h(p\eta_1)\, h^*(p\eta_2) \, n_p(\eta) + h(p\eta_1)\, h(p\eta_2) \,\kappa_p(\eta) + c.c. \right], \nn \\ {\rm where} \quad
n_p(\eta) \approx \frac{\l^2 \, S_{D-2}}{(2\pi)^{D-1}} \, \int d^{D-1}q \iint_{\eta_0}^\eta d\eta_3 \, d\eta_4 \, (\eta_3\, \eta_4)^{\frac{D-3}{2}} \times \nn \\ \times h^*\left(p\eta_3\right) \, h\left(p\eta_4\right) \, h^*(q\,\eta_3)\,  h(q\eta_4) \, h^*(|q-p|\eta_3)\, h(|p-q|\eta_4), \nn \\ {\rm and} \quad
\kappa_p(\eta) \approx - \frac{2\,\l^2 \, S_{D-2}}{(2\pi)^{D-1}} \, \int d^{D-1}q \int_{\eta_0}^\eta d\eta_3 \int_{\eta_0}^{\eta_3} d\eta_4 \, (\eta_3\, \eta_4)^{\frac{D-3}{2}} \times \nn \\ \times h^*\left(p\eta_3\right) \, h^*\left(p\eta_4\right) \, h^*(q\eta_3)\,  h(q\eta_4) \, h^*(|q-p|\eta_3)\, h(|q-p|\eta_4).
\eqa
The details of a similar calculation in EPP can be found in \cite{Krotov:2010ma,Akhmedov:2011pj,Akhmedov:2012pa}.
The largest IR contribution to (\ref{CPP}) comes from the region of integration over $dq$, where $q\gg p$. It is instructive to study the character of such divergences for different choices of the harmonics $h(p\eta)$.
If $h(x)\propto {\cal H}^{(2)}_{i\mu} (x)$ the IR divergence is present both in $n_p$ and $\kappa_p$ and is proportional to $\log\left(\eta/\eta_0\right) = t - t_0$, if $p\eta \ll 1$, which is the proper time elapsed from $\eta_0 = e^{t_0}$ to $\eta = e^t$. If, however, $p\eta \gg 1$, then the divergence is proportional to $\log\left(1/p\eta_0\right)$. The explanation for the latter behavior is given at the end of the next subsection.

The in-harmonics of the CPP are proportional to the Bessel function $Y_{i\mu}(x)$.
In this case $n_p(\eta)$ diverges as $\log\left(\eta/\eta_0\right)$, if $p\eta \ll 1$, and as $\log\left(1/p\eta_0\right)$, when $p\eta \gg 1$. At the same time, $\kappa_p(\eta)$ does not have any IR divergence. We interpret this fact as that the backreaction, due to the particle production on the background state,
corresponding to the in--harmonics, is small. That is the reason why we choose these harmonics for the proper definition of the quasi--particles in the IR limit \cite{Akhmedov:2011pj} (see also the discussion below). For any other type of $\alpha$-harmonics the IR behavior of $n_p$ and $\kappa_p$ is similar to that of ${\cal H}^{(2)}_{i\mu} (p\eta)$ harmonics.

\subsection{The physical interpretation of the one-loop IR divergence}

Now we are going to show that
$n_p$ given in (\ref{CPP}) is the total particle number density per comoving volume, $\langle a^+_p \, a
_p \rangle$, created by the gravitational field during the time passed from $\eta_0$ to $\eta$.
Similarly $\kappa_p$ is the measure of the strength of the backreaction, $\langle a_p \, a_{-p} \rangle$, due to the particle production process on the background state during the same time period.

To explain the statement of the last paragraph, let us forget for the moment about the presence of $\kappa_p$ and consider the situation in flat space with the plane wave harmonics. If the time of the turning on selfinteractions is $t_0$ and the time of the observation is $t$, then the kinetic equation is (the derivation can be found e.g. in \cite{Akhmedov:2011pj}):

\bqa\label{coll}
\frac{dn_p}{dt} \propto - \lambda^2 \, \int\frac{d^{D-1}q}{\epsilon(p)\,\epsilon(q)\, \epsilon(p-q)} \times \nonumber \\  \left\{ \int_{t_0}^t dt' \cos\left[\left(-\epsilon(p) + \epsilon(q) + \epsilon(p-q)\right)\, (t-t')\right] \, \left[(1+n_p)\, n_{q}\, n_{|p-q|}^{\phantom{\frac12}} - n_p \, (1+n_{q})\, (1+n_{|p-q|})\right](t) \right. \nonumber \\ + 2\, \int_{t_0}^t dt' \cos\left[\left(- \epsilon(q) + \epsilon(q-p) + \epsilon(p)\right)\, (t-t')\right] \, \left[n_{q}\,(1 + n_{|q-p|})\, (1 + n_p)^{\phantom{\frac12}} - (1+n_{q})\, n_{|q-p|} \, n_p\right](t) \nonumber \\ +
\left. \int_{t_0}^t dt' \cos\left[\left(\epsilon(q) + \epsilon(q+p) + \epsilon(p)\right)\, (t-t') \right] \, \left[(1 + n_{q})\,(1 + n_{q+p})\, (1 + n_p)^{\phantom{\frac12}} - n_{q}\, n_{q+p} \, n_p\right](t) \right\},
\eqa
where $\epsilon(p)$ is the energy of the particle with the momentum $\vec{p}$. The RHS of this equation is the collision integral $I_{Coll}$. In the limit $(t-t_0)\to \infty$ the $dt'$ integrals transform into the $\delta$-functions ensuring energy conservation in some processes, which we are going to define now.

The expressions multiplying the three $dt'$ integrals in (\ref{coll}) have the following meaning. The first term describes the competition between the following two processes. One of them is that the particle $\vec{p}$ decays into two excitations --- $\vec{q}$ and $\vec{p}- \vec{q}$. It corresponds to the term $n_p \, (1+n_{q})\, (1+n_{|p-q|})$ and appears with the minus sign in the collision integral, because it describes the loss of the excitation with the momentum $\vec{p}$. The inverse gain process, corresponding to the term $(1+n_p)\, n_{q}\, n_{|p-q|}$ with the plus sign, is such that two particles, $\vec{q}$ and $\vec{p}-\vec{q}$, are merged together to create the excitation $\vec{p}$.

The second term in (\ref{coll}) also describes two competing processes. The first of them is such that the particle $\vec{p}$ joins together with another one, $\vec{q}-\vec{p}$, to create the excitation with the momentum $\vec{q}$. This is the loss process. The inverse gain process is the one in which the particle $\vec{q}$ decays into two, one of which is with the momentum $\vec{p}$. The coefficient 2 in front of this term is just the combinatoric factor.

Similarly the third term describes two processes.
The gain process is when three particles, one of which is with the momentum $\vec{p}$,
are created by an external field, if any. The loss process is when three such excitations are annihilated into the vacuum.

All these six types of processes are not allowed by the energy--momentum conservation for {\it massive}
fields with $\phi^3$ selfinteraction in {\it flat} space-time. Hence, the collision integral is vanishing as $(t-t_0)\to\infty$.

However, in dS space they are allowed \cite{Myrhvold}--\cite{Volovik:2008ww}, because there is no energy conservation. As we show below the collision integral in dS space can be obtained from (\ref{coll}) as follows. One exchanges $dt'$ for $d\eta'$, multiplies by the proper weight coming from the metric, and uses under the corresponding integrals the dS harmonics, $g_p(\eta) = \eta^{\frac{D-1}{2}}\, h(p\eta)$, instead of the plane waves, $e^{-i\, \epsilon(p) \, t}/\sqrt{\epsilon(p)}$. Then, in place of the above mentioned $\delta$-functions, which ensure energy conservation, there are some expressions whose physical meaning is that they define the differential rates (per given range of $\vec{q}$) of the processes described in the above three paragraphs.

Now put $n_p$ to zero under the collision integral (\ref{coll}) with the substitutions described in the previous paragraph. The only term which survives is present in the last line of (\ref{coll}) and describes particle creation by the gravitational field. If one takes the integral of this term over the conformal time from $\eta_0$ to $\eta$ he obtains exactly the expression for $n_p(\eta)$ represented in (\ref{CPP}).
Using the collision integral for $\kappa_p$, derived in \cite{Akhmedov:2011pj,Akhmedov:2012pa} (see also the discussion below), one can make similar observation about the origin of the corresponding expression in (\ref{CPP}).

Thus, in the non-stationary situations, when $I_{Coll}$ is not zero, it is natural to expect the linear divergence, $\propto (t-t_0)$, in $n_p$ and $\kappa_p$. That is, indeed, should be true for the homogeneous in time background fields (such as e.g. constant electric fields in QED), when particle production rates are constant in time.
But, in time dependent metric of dS space one encounters a bit different situation.
In particular, in EPP the largest IR contribution has the form of $\lambda^2 \,\log(p\eta)$ and there is no IR divergence as one takes the moment of turning on selfinteractions to past infinity of the EPP ($\eta_0 \to + \infty$). That happens because the creation of particles with comoving momentum $p$ effectively starts at $\eta_* \sim 1/p$ rather than at $\eta_0\to +\infty$. Similarly in the case of CPP the particle creation process goes on until $\eta_* \sim 1/p$ and then stops. We encounter such a situation, because,
first of all, the collision integral is not just a constant and depends on time. Second, past (future) infinity in EPP (CPP) corresponds to the UV limit of the physical momentum. In this limit $g_p(\eta)$ harmonics behave as plain waves $e^{i\,p\,\eta}$. As the result all the rates inside the collision integral can be well approximated by $\delta$-functions ensuring energy conservation \cite{Akhmedov:2011pj}. The latter fact has a clear physical meaning, because the modes whose wavelength is much smaller than the cosmological horizon size, practically do not feel the curvature of the space--time. Hence, for them the conditions should be similar to those in flat space.

Thus, the collision integral becomes negligible as $p\eta \to \infty$ both in EPP and CPP. That is the reason why in the EPP, independently of the type of the harmonics, one can put the moment of turning on selfinteractions, $\eta_0$, to past infinity. Similarly these observations explain the $\log(1/p\eta_0)$ behavior of the leading IR contribution in the CPP, when $\eta > 1/p$.

\subsection{Kinetic equation in the contracting Poincare patch}

In (\ref{CPP}) we have found the created particle density $n_p$ and anomalous average $\kappa_p$ under the implicit assumption that throughout the time period $[\eta_0, \eta]$ both $n_p$ and $\kappa_p$ have been zero inside the collision integral. In fact, we have been using the tree-level propagators in the calculation of (\ref{CPP}).
For the proper consideration of the phenomenon one has to redo the calculation and take into account the change of $n_p$ and $\kappa_p$ in time. If we care only about IR contributions this is done with the use of the following ansatz for the IR behavior of the exact Keldysh propagator \cite{Akhmedov:2012pa}:

\bqa\label{anzatz}
D_{p}^{K}(\eta_1,\eta_2) = (\eta_1\eta_2)^{\frac{D-1}{2}}\left[h(p\eta_1)\, h^*(p\eta_2) \, n_p(\eta) + h(p\eta_1)\, h(p\eta_2) \,\kappa_p(\eta) + c.c. \right], \quad \eta = \sqrt{\eta_1\eta_2},
\eqa
where one has to assume the usual separation of scales in the kinetic problem, i.e. that $n_p$ and $\kappa_p$ are slow functions of $\eta$ in comparison with the harmonics $h(p\eta)$. Such a separation of scales is quite obvious for the massive fields from the principal series, because their harmonics do oscillate in the IR limit. However, for the light fields from the complementary series the separation of scales is not so transparent, because their harmonics do not oscillate in the same limit.

It is not hard to see that after the assumption about the IR behavior of the exact propagator (\ref{anzatz}), the problem of the calculation of $n_p$ and $\kappa_p$ reduces to the solution of the Dyson-Schwinger equation in the Schwinger-Keldysh diagrammatic technique \cite{Akhmedov:2012pa}. In the IR limit, which is of our interest, the Dyson--Schwinger equation reduces to a variant of the kinetic equation.
One can write the kinetic equation both in EPP and CPP \cite{Akhmedov:2011pj,Akhmedov:2012pa}. With the use of the following matrixes:

\bqa
N_p(\eta_1,\, \eta_2) = \eta_2^{\frac{D-1}{2}} \, \left(
  \begin{array}{cc}
    n_p(\eta_1) \, h^*(p\eta_2) & \kappa_p(\eta_1) \, h(p\eta_2) \\
    \kappa_p(\eta_1) \, h(p\eta_2) & n_p(\eta_1) \, h^*(p\eta_2) \\
  \end{array}
\right), \quad P = \left(
  \begin{array}{cc}
    0 & 1 \\
    1 & 0 \\
  \end{array}
\right)
\eqa
the system of kinetic equations for $n_p$ and $\kappa_p$ can be written compactly. The real equation has the form:

\bqa\label{callintPP0}
\frac{d n_p(\eta)}{d\eta} = 2\,\lambda^2 \, \int \frac{d^{D-1}q}{(2\pi)^{(D-1)}} \, \int_{0}^{\infty} \frac{d\eta'}{\left(\eta\, \eta'\right)^D} \, \, {\rm Re}\, \, Tr \nonumber \\ \left\{C_{p,q,p-q}(\eta)\, \left[\left(1+N^*_p\right)\, N_{q}\, N_{p-q}^{\phantom{\frac12}} - \,\, N^*_p \, \left(1+N_{q}\right)\, \left(1+N_{p-q}\right)\right](\eta,\eta') +  \right. \nonumber \\ + 2 \, C_{q,q-p,p}(\eta)\, \left[ N^*_{q}\, \left(1+N_{q-p}\right)\,\left(1+N_p\right)^{\phantom{\frac12}} - \,\,\left(1+N^*_{q}\right)\, N_{q-p}\, N_p \right](\eta,\eta') + \nonumber \\ \left.
+ D_{p,q,p+q}(\eta)\,\left[ \left(1+N_{p}\right)\, \left(1+N_{q}\right) \,\left(1+N_{p+q}\right)^{\phantom{\frac12}} - \,\, N_{p}\, N_{q}\, N_{p+q} \right](\eta,\eta')^{\phantom{\frac12}}\right\} + \left[N\to P\,N\right].
\eqa
At the same time the complex equation is as follows:

\bqa\label{callintPP00}
\frac{d \kappa_p(\eta)}{d\eta} = - 2\,\lambda^2 \, \int \frac{d^{D-1}q}{(2\pi)^{(D-1)}} \, \int_{0}^{\eta} \frac{d\eta'}{\left(\eta\, \eta'\right)^D} \, \, \left(\vec{p}\to - \vec{p}\right)\, \, Tr \nonumber \\ \left\{C_{p,q,p-q}(\eta)\,\left[ \left(1+N_{p}\right)\, \left(1+N_{q}\right) \,\left(1+N_{p-q}\right)^{\phantom{\frac12}} - \,\, N_{p}\, N_{q}\, N_{p-q} \right](\eta,\eta') +  \right. \nonumber \\ + 2\,C^*_{q,q-p,p}(\eta)\, \left[ N^*_{q}\, \left(1+N_{q-p}\right)\,\left(1+N_p\right)^{\phantom{\frac12}} - \,\,\left(1+N^*_{q}\right)\, N_{q-p}\, N_p \right](\eta,\eta') + \nonumber \\ \left.
+ D^*_{p,q,p+q}(\eta)\, \left[\left(1+N_p\right)\, N^*_{q}\, N_{p+q}^{*\phantom{\frac12}} - \,\, N_p \, \left(1+N^*_{q}\right)\, \left(1+N^*_{p+q}\right)\right](\eta,\eta')^{\phantom{\frac12}}\right\} - \left[N\to P\,N\right].
\eqa
The term $\left[N\to P\,N\right]$ means that we have to add to the explicitly written expressions in (\ref{callintPP0}) and (\ref{callintPP00}) the same quantities with every $N$ substituted by the product $P\,N$; Re $Tr$ means that one has to take the real part and the trace of the expression following after these signs; $\left(\vec{p}\to - \vec{p}\right)\,Tr$ means that one has to take the trace and add to the expression following after these signs the same term with the exchange $\vec{p}\to - \vec{p}$. Finally $C_{k_1k_2k_3}(\eta) = \eta^{\frac{3\, (D-1)}{2}}\,h^*(k_1\eta)\,h(k_2\eta)\, h(k_3\eta)$ and $D_{k_1k_2k_3}(\eta) = \eta^{\frac{3(D-1)}{2}}\,h(k_1\eta)\,h(k_2\eta)\, h(k_3\eta)$.

Actually the system (\ref{callintPP0}) and (\ref{callintPP00}) represents just a preliminary version of the kinetic equations. It is not yet suitable to sum only leading IR terms in all loops. It contains the latter contributions, but on top of that it also accounts for some of the subleading expressions. In particular, one should restrict the $dq$ integrals in (\ref{callintPP0}) and (\ref{callintPP00}) to the region $q \gg p$. It is this region from where the leading IR contribution does come. For more details see \cite{Akhmedov:2011pj,Akhmedov:2012pa} (see also the discussion in the next subsection).

Due to the presence of $\kappa_p$, in (\ref{callintPP0}) we have extra terms on top of those which are shown in (\ref{coll}). All of them can be obtained from (\ref{coll}) via the simultaneous substitutions of $(1+n_{k})$'s and $n_{k}$'s by $\kappa_{k}$'s or $\kappa^*_{k}$'s. E.g. we encounter terms of the type:

\bqa\label{9}
\left[(1+n_p)\, \kappa_{k}\, n_{p-k} - n_p \, \kappa_{k}\, (1+n_{p-k})\right].
\eqa
The meaning of (\ref{9}) is that it describes the following two competing processes --- the particle with the momentum $\vec{p}$ is lost (gained) in such a situation, in which instead of the creation (annihilation) of the two particles, $\vec{k}$ and $\vec{p}-\vec{k}$, we obtain single $\vec{p}-\vec{k}$ excitation and the missing momentum $\vec{k}$ is gone into (taken from) the background quantum state of the theory.

Similarly we encounter terms of the type

\bqa
\left[(1+n_p)\, \kappa_{k}\, \kappa_{p-k} - n_p \, \kappa_{k}\, \kappa_{p-k}\right]
\eqa
which describe the processes in which both momenta $\vec{k}$ and $\vec{p}-\vec{k}$ are coming from (going to) the background state. These observations, in particular, justify the interpretation of $\kappa_p$ as the measure of the strength of the backreaction on the background state of the theory.

\subsection{Solving the system of kinetic equations}

Spatially homogeneous kinetic equations in EPP and CPP can be trivially related to each other, because to perform the map from EPP to CPP one just has to flip the limits of $d\eta'$ integration inside the collision integrals in (\ref{callintPP0}) and (\ref{callintPP00}). On top of that it is necessary to make the change of $h(x)$ to $h^*(x)$, because the positive energy harmonics are exchanged with the negative ones under the flip of the time direction.
As the result, solutions in both patches also can be mapped to each other. That is physically meaningful because the system of kinetic equations (\ref{callintPP0}) and (\ref{callintPP00}) is valid for the spatially homogeneous states and is imposed on the quantities which are attributed to the comoving volume.

Before continuing with the solution of (\ref{callintPP0}) and (\ref{callintPP00}) let us stress that these equations are not yet unambiguously defined, because one has to specify harmonics $h(x)$ in their collision integrals. This issue is related to the question of the proper definition of quasi-particles in the IR limit. One should keep in mind, however, that the quantum kinetic (or better to say Dyson-Schwinger) equations and their solutions for different choices of $h(p\eta)$ can be mapped to each other via Bogolubov transformations.

Intuition from condensed matter should tell one that the proper choice of $h(p\eta)$ has to correspond to the situation in which there is a solution with $\kappa_p = 0$.
It was shown in \cite{Akhmedov:2011pj} that in EPP there is such a solution of the IR limit of (\ref{callintPP0}) and (\ref{callintPP00}), if one chooses out-Jost harmonics, $h(x) \sim J_{i\mu}(x)$. Moreover it is stable under small linearized perturbation of $\kappa_p$ \cite{Akhmedov:2012pa}. The systems of kinetic equations for the other harmonics do not allow such a solution.

All these observations can be extended to the CPP. Thus, for the out-Jost (in-Jost) harmonics in EPP (CPP) the system (\ref{callintPP0}) and (\ref{callintPP00}) can be reduced to the single equation (\ref{callintPP0}), in which $\kappa_p$ is set to zero \cite{Akhmedov:2012pa}\cite{Akhmedov:2011pj}:

\bqa\label{callintPP2}
\frac{d n_p(\eta)}{d\log \eta} = \frac{\lambda^2 \, S_{D-2}}{(2\pi)^{D-1}\, \mu} \, \int_0^{\infty} dq \, \eta \,\left(q \eta\right)^{\frac{D-1}{2}} \, \int^{\infty}_{0} d\eta' \, q \, \left(q \eta'\right)^{\frac{D-1}{2}} \times \nonumber \\ \times \Biggl\{ {\rm Re} \left[(q\eta)^{-i\mu} \, V(q\eta) \,\left(q\eta'\right)^{i\mu}\, V^*(q\eta')\right] \, \left[(1+n_p)\, n_q^{2\phantom{\frac12}} - \,\, n_p \, (1+n_q)^2\right](\eta) \Biggr. + \nonumber \\
+ 2\,{\rm Re} \left[(q\eta)^{i\mu} \, W(q\eta) \, \left(q\eta'\right)^{- i\mu} \, W(q\eta')\right] \, \left[ n_q\, (1+n_q)\,(1+n_p)^{\phantom{\frac12}} - \,\,(1+n_q)\, n_q\, n_p \right](\eta) + \nonumber \\
+ \Biggl. {\rm Re} \left[(q\eta)^{i\mu} \, V(q\eta) \, \left(q\eta'\right)^{- i\mu} \, V^*(q\eta')\right] \, \left[ (1+n_q)^2 \,(1+n_p)^{\phantom{\frac12}} - \,\, n_q^2\, n_p \right]^{\phantom{2}}(\eta)\Biggr\}.
\eqa
To obtain (\ref{callintPP2}) from (\ref{callintPP0}) we have neglected $p$ with respect to $q$ under the integral and Taylor expanded $h(p\eta)$ and $h(p\eta')$ in the limit $p\eta, p\eta'\to 0$. The latter approximations allow one to obtain the equation which accounts for only leading IR contributions in all loops \cite{Akhmedov:2011pj}. Furthermore, we have subtracted from $h^2(x)$, $\left(h^*(x)\right)^2$ and $|h(x)|^2$ some polynomials of $1/|x|$ to obtain $V(x) = h^2 (x) - \frac{const}{|x|} - \dots$ and $W(x) = \left|h(x)\right|^2 - \frac{const}{|x|} - \dots$. That makes the collision integral well defined and does not change the IR behavior \cite{Akhmedov:2011pj}.

Eq. (\ref{callintPP2}) does not have the Plankian distribution as its stationary solution, because there is no energy conservation. However, one can find a distribution which annihilates the collision integral of (\ref{callintPP2}) in a different manner: while Plankian distribution in ordinary kinetic equation annihilates each term, describing aforementioned couples of competing processes, separately, in the present case we will see the cancelation between different terms. The situation is somewhat similar to the one in which appears the stationary Kolmogorov scaling solution in the turbulent hydrodynamics, as is natural to expect in a system with energy pumping.

For very small densities the only two terms, which survive in the collision integral of (\ref{callintPP2}) are those, which are responsible for the particle production from the vacuum and for the decay of the produced particles into two \cite{Akhmedov:2011pj}. As the result in the EPP the behavior of the solution in the vicinity of the stationary distribution (i.e. as $p\eta \ll 1$) is as follows \cite{Akhmedov:2011pj}: $n_p(\eta) = \frac{\Gamma_2}{\Gamma_1} \left[C\, \left(p\eta\right)^{\Gamma_1} + 1\right]$, where $C$ is some real constant which depends on the initial conditions; $\Gamma_2$ is the total particle production rate and $\Gamma_1$ is the total particle decay rate. This solution is self consistent only for heavy enough particles $m\gg 1$, because then the equilibrium distribution, $\Gamma_2/\Gamma_1 \approx e^{-2\pi \mu}$, is much less than unity.

Such a EPP's stationary solution is mapped to

\bqa\label{stage}
n_p(\eta) = \frac{\Gamma_2}{\Gamma_1} \left[C\, \left(\frac{\eta_0}{\eta}\right)^{\Gamma_1} + 1\right],
\eqa
after the conversion into the CPP. For the given comoving momentum $p$ it is valid until $\eta_* \sim 1/p$.
As $p\eta$ becomes comparable to unity, the approximation, in which (\ref{callintPP2}) is derived, brakes down.
But as we have explained above, right after $\eta_*$ the particle production and decay processes stop, i.e. $n_p$ remains constant. One encounters a similar situation in EPP: before $\eta_*\sim 1/p$ we can not use (\ref{callintPP2}), but also before this moment of time the particle production is suppressed.

In (\ref{stage}) we see the example of the restoration of dS isometry in CPP after the summation of all leading loop contributions. The question is under what type of initial conditions the system in question drops into such a stage as (\ref{stage}) in CPP (and/or in EPP). To obtain the answer on this question one has to perform the Bogolubov rotation to the out (in) harmonics in EPP (CPP) and check if the resulting values of $n_p$ and $\kappa_p$ are small or not, right at the moment when they start to change (which is $\eta_0$ in CPP and $\eta_*$ in EPP). If they are small enough, then it is natural to expect that $\kappa_p$ flows to zero and $n_p$ flows to Gibbons--Hawking value in the IR limit.

Note that the number density per physical volume, which scales as $1/\eta^{D-1}$, decays in EPP and grows in CPP. Hence, it may seem that the backreaction from the produced particles eventually will become relevant in CPP.
That is indeed true in the case of the presence of any slight spatially inhomogeneous perturbation.
But in the circumstances of the ideal spatial homogeneity that is not true, because the presence of the stationary state means that the dS isometry is restored. As the result, in terms of the invariant expressions, such as physical volume and physical momentum $p\eta$, all invariant quantities remain finite. E.g. the total particle density per physical volume is equal to $N = \eta^{D-1}\,\int d^{D-1} p\, n_p(\eta)$ and does not explode.

However, we are going to see now that the kinetic equation in question has yet another peculiar solution both in CPP and EPP. In \cite{Akhmedov:2011pj} it was shown that in CPP there is a solution as follows: $n(\eta) \sim \frac{1}{A - \bar{\Gamma}\, \log\eta} \sim \frac{1}{\bar{\Gamma}\,\log\frac{\eta_*}{\eta}}$,
for very low momenta, $p\eta \to 0$. It is valid for $\eta < \eta_* = e^{const/\lambda^2}\gg 1$; $A$ is a constant which depends on the initial state, and $\bar{\Gamma} \propto \frac{\lambda^2}{m^2}$. The interpretation of this solution is as follows: Due to the production of particles at low momenta (both by the gravitational field and by the decays of particles from higher levels) at high enough initial densities, sooner or latter we have the explosion of their density per comoving volume.

This solution is mapped to

\bqa\label{dyuzh}
n(\eta) \sim \frac{1}{\bar{\Gamma} \, \log\frac{\eta}{\eta_*}}
\eqa
in the EPP. It is also valid for very low momenta, $p\eta\to 0$, for $\eta > \eta_*$, where $\eta_*= e^{-const/\lambda^2}\ll 1$.

It is worth stressing at this moment that the solution (\ref{dyuzh}) is possible because of two specific features of dS space \cite{Akhmedov:2011pj}. The first one is the flatness of the IR spectrum in dS space --- weak dependence of the harmonics on the momenta in the IR limit. And the second feature is that background gravitational field makes possible particle creation and decay processes. In particular, one can show that the scattering processes in $\phi^4$ theory, which are allowed even in flat space, do not contribute to the analog of (\ref{dyuzh}): only the particle production and decay processes do contribute.

Thus, even in the EPP we have the explosive growth of particle density for big enough its initial values. The explosion happens at finite proper time and expansion of EPP will not save the situation. That will cause a modification of the dS metric.

Here also it is appropriate to ask the question: under what initial conditions the system drops into such an explosive stage both in EPP and CPP? We are not yet in a position to give a conclusive answer on this question.

\section{Global de Sitter space-time}

To begin with we would like to show, following \cite{Krotov:2010ma}, that in global dS the moment of turning on selfinteractions cannot be taken to past infinity. To find the IR divergences at one loop, we can use the metric of either fig. 1 or fig. 2. However, to define the kinetic equation in global dS we prefer to work with the slicing depicted in fig. 2. Because, as we explain above, for the metric of fig. 1 the collision integral vanishes for the times $\eta_- > 1/p$ and $\eta_+ > 1/p$, which confuses the whole picture of particle kinetics.

Unlike higher dimensional dS with the spatial slicing of fig. 2, in 2D global dS space the one-loop calculation can be made quite explicit. So let us present it before continuing with the IR divergences in general dimensions.
Below we are going to work with even space-time dimensional dS space. In odd dimensional dS space the situation is different, but in many respects is similar.

Due to spatial homogeneity of 2D global dS space together with the initial state, we find it convenient to perform the Fourier transform of all quantities along the spatial direction $\varphi \in [0,2\pi]$: $D^{K,R,A}_p(t_1, t_2) \equiv \int_0^{2\pi} d\varphi \, e^{i\, p \, \varphi} D^{K,R,A}(t_1, \varphi; t_2, 0)$. Then the one-loop contribution to $D^K_p$ is as follows:

\bqa\label{DSDK}
D_{1p}^{K}(t_1,t_2) = - \l^2 \sum_{q=-\infty}^{+\infty} \iint_{t_0}^{+\infty} dt_3 dt_4 \cosh(t_3)\, \cosh(t_4) \, \Biggl[ \Biggr. D_{0p}^R(t_1,t_3) \, D_{0q}^K(t_3,t_4) \, D_{0(p-q)}^K(t_3,t_4) \, D_{0p}^A(t_4,t_2) + \nn\\ + 2 \, D_{0p}^R(t_1,t_3) \, D_{0q}^R(t_3,t_4) \, D_{0(p-q)}^K(t_3,t_4) \, D_{0p}^K(t_4,t_2) + 2 \, D_{0p}^K(t_1,t_3) \, D_{0q}^K(t_3,t_4) \, D_{0(p-q)}^A(t_3,t_4)\, D_{0p}^A(t_4,t_2) + \nn\\ - \fr{1}{4} \, D_{0p}^R(t_1,t_3) \, D_{0q}^R(t_3,t_4) \, D_{0(p-q)}^R(t_3,t_4) \, D_{0p}^A(t_4,t_2) -\fr{1}{4} \, D_{0p}^R(t_1,t_3) \, D_{0q}^A(t_3,t_4) \, D_{0(p-q)}^A(t_3,t_4) \, D_{0p}^A(t_4,t_2) \Biggl. \Biggr].
\eqa
Here the Keldysh propagator is $D_{0p}^{K}(t_1,t_2) = {\rm Re}\left[g_p(t_1)g^*_p(t_2)\right]$, retarded and advanced propagators are $D_{0p}^{R}(t_1,t_2) = \theta\left(t_1-t_2\right) \,2\,{\rm Im}\left[g_p(t_1)g^*_p(t_2)\right]$ and $D_p^{A}(t_1,t_2) = - \theta\left(t_2-t_1\right) \, 2\,{\rm Im}\left[g_p(t_1)g^*_p(t_2)\right]$; $g_p(t)$ is a solution of the Klein--Gordon equation, which we do not specify so far; $t_0$ is the moment when selfinteractions are turned on.

The above one-loop expression can be rewritten as:

\bqa
D_{1p}^{K}(t_1,t_2) = -\frac{\l^2}{2} \, g_p(t_1)\, g^*_p(t_2) \, \sum_q \iint_{t_0}^{+\infty} dt_3 dt_4 \cosh(t_3)\, \cosh(t_4) g_p^*(t_3)\, g_p(t_4)\times \nn \\ \times \Biggl\{ \Biggr. - \theta(t_1 - t_3)\, \theta(t_2 - t_4) \, \Biggl[g_q(t_3) \, g_q^*(t_4) \, g_{|p-q|}(t_3)\,g_{|p-q|}^*(t_4) + c.c.\Biggr] + \nn \\ \Biggl[ \theta(t_1 - t_3)\, \theta(t_3 - t_4) + \theta(t_2 - t_4)\, \theta(t_4 - t_3)\Biggr] \, \Biggl[g_q(t_3) \, g_q^*(t_4) \, g_{|p-q|}(t_3)\,g_{|p-q|}^*(t_4) - c.c.\Biggr] \Biggl. \Biggr\} - \nn \\
-\frac{\l^2}{2} \, g_p(t_1)\, g_p(t_2) \, \sum_q \iint_{t_0}^{+\infty} dt_3 dt_4 \cosh(t_3)\, \cosh(t_4) g_p^*(t_3)\, g^*_p(t_4) \times \nn \\ \times \Biggl\{ \Biggr. \theta(t_1 - t_3)\, \theta(t_2 - t_4) \, \Biggl[g_q(t_3) \, g_q^*(t_4) \, g_{|p-q|}(t_3)\,g_{|p-q|}^*(t_4) + c.c.\Biggr] + \nn \\ \Biggl[ \theta(t_1 - t_3)\, \theta(t_3 - t_4) - \theta(t_2 - t_4)\, \theta(t_4 - t_3)\Biggr] \, \Biggl[g_q(t_3) \, g_q^*(t_4) \, g_{|p-q|}(t_3)\,g_{|p-q|}^*(t_4) - c.c.\Biggr] \Biggl. \Biggr\} + c.c.
\eqa
We need to find the leading divergent contribution to this quantity in the limit $t\equiv \frac{t_1 + t_2}{2}\to +\infty$, $\tau \equiv t_1 - t_2 = const$ and $t_0 \to -\infty$. First, to do that we can substitute $t_{1,2}$ by $t$ in the arguments of the $\theta$-functions. Second, if $t\to+\infty$ and $t_0 \to -\infty$, then the leading IR contributions to $D_{1p}^K$ can appear only from the regions of integration over $t_{3,4}$ in the vicinity of $t$ and $t_0$.

As we have explained in the introductory section, for arbitrary dimension in these regions $g_p(t)$ can be approximated by

\bqa
g_p(t) \approx \left\{
                  \begin{array}{cc}
                    \eta_+^{\frac{D-1}{2}} \, h_+(p\eta_+), & \eta_+ = e^{-t}, \quad t \to +\infty \\
                    \eta_-^{\frac{D-1}{2}} \, h_-(p\eta_-), & \eta_- = e^t, \quad t\to -\infty \\
                  \end{array}
                \right.
\eqa
for the large enough $p$.
Then as $p\eta\to 0$ and $p\eta_0\to 0$ the leading IR contribution to $D_{1p}^K$ can be estimated in any dimension. It is contained in the expressions as follows \cite{Krotov:2010ma}:

\bqa\label{Leadingoneloop}
D_{1p}^{K}(t_1,t_2) = g_p(t_1)\, g^*_p(t_2) \, n_p(t) + g_p(t_1)\, g_p(t_2) \,\kappa_p(t) + c.c., \nn \\
{\rm where} \quad n_p(t) \approx \frac{\l^2 \, S_{D-2}}{(2\pi)^{D-1}} \, \int d^{D-1}q \iint_\infty^\eta d\eta_3 \, d\eta_4 \, (\eta_3\, \eta_4)^{\frac{D-3}{2}} \times \nn \\ \times h_+^*\left(p\eta_3\right) \, h_+\left(p\eta_4\right) \, h_+^*(q\,\eta_3)\,  h_+(q\eta_4) \, h^*_+(|q-p|\eta_3)\, h_+(|p-q|\eta_4)  + \nn \\
+ \frac{\l^2 \, S_{D-2}}{(2\pi)^{D-1}} \, \int d^{D-1}q \iint^\infty_{\eta_0} d\eta_3 \,\eta_4\, (\eta_3\, \eta_4)^{\frac{D-3}{2}} \times \nn \\ \times h_-^*\left(p\eta_3\right) \, h_-\left(p\eta_4\right) \, h_-^*(q\eta_3)\,  h_-(q\eta_4)\, h_-^*(|q-p|\eta_3) \, h_-(|q-p|\eta_4), \nn \\ {\rm and} \quad
\kappa_p(t) \approx - \frac{2\,\l^2 \, S_{D-2}}{(2\pi)^{D-1}} \, \int d^{D-1}q \int_\infty^\eta d\eta_3 \int_{\infty}^{\eta_3} d\eta_4 \, (\eta_3\, \eta_4)^{\frac{D-3}{2}} \times \nn \\ \times h_+^*\left(p\eta_3\right) \, h_+^*\left(p\eta_4\right) \, h_+^*(q\eta_3)\,  h_+(q\eta_4) \, h_+^*(|q-p|\eta_3)\, h_+(|q-p|\eta_4) - \nn \\
- \frac{2\,\l^2 \, S_{D-2}}{(2\pi)^{D-1}} \, \int d^{D-1}q \int^\infty_{\eta_0} d\eta_3 \int^{\eta_3}_{\eta_0} d\eta_4 \, (\eta_3\, \eta_4)^{\frac{D-3}{2}} \times \nn \\ \times h_-^*\left(p\eta_3\right) \, h_-^*\left(p\eta_4\right) \, h_-^*(q\eta_3) \,  h_-(q\eta_4)\, h^*_-(|q-p|\eta_3) \, h_-(|q-p|\eta_4).
\eqa
Here $\eta = e^{-t}$ and $\eta_0 = e^{t_0}$. Thus, IR divergence in global dS is just the sum of the contributions from the contracting (past infinity) and expanding (future infinity) regions (Poincare patches) of the whole space.

Obviously one can do the same calculation also with the metric of fig. 1 and obtain the same as (\ref{Leadingoneloop}) result for the IR divergences, when $t\to +\infty$ and $t_0\to-\infty$.

\subsection{IR divergence for different harmonics}

In this subsection we describe the character of the IR divergences for different $h(p\eta)$ --- the solutions of the Bessel equation. That could help to specify the convenient choice of the harmonics for the proper definition of quasi--particles in the IR limit.

We start with the Euclidian harmonics \cite{Mottola:1984ar},\cite{Allen:1985ux} which are CPT invariant. I.e. in this case, $h_-^*(x) = h_+(x) \propto {\cal H}^{(1)}_{i\mu}(x)$, where ${\cal H}^{(1)}_{i\mu}$ is the Hankel function of the first kind. Using that $h_+(x) \approx A_+ \, x^{i\mu} + A_-\, x^{-i\mu}$ as $x\to 0$ (with some known complex constants $A_\pm$), we obtain the IR divergence as follows:

\bqa
n_p(t) \approx \frac{\l^2 S_{D-2}}{(2\pi)^{D-1}}\, \log\left(\frac{1}{p^2 \, \eta \, \eta_0}\right)\, \iint_0^\infty dx_3 \, dx_4 \, \left(x_3 \, x_4\right)^{\frac{D-3}{2}}\times \nn \\ \times \left[\left|A_+\right|^2 \, \left(\frac{x_4}{x_3}\right)^{i\mu} + \left|A_-\right|^2 \, \left(\frac{x_3}{x_4}\right)^{i\mu}\right]\, h^2(x_3)\, \left[h^*(x_4)\right]^2, \nn \\
\kappa_p(t) \approx - \frac{2\,\l^2 S_{D-2}}{(2\pi)^{D-1}}\, \log\left(\frac{1}{p^2 \, \eta \, \eta_0}\right)\, A_+ \, A_- \, \int^0_\infty dx_3 \,\int_\infty^{x_3} dx_4 \, \left(x_3 \, x_4\right)^{\frac{D-3}{2}}\times \nn \\ \times \left[\left(\frac{x_4}{x_3}\right)^{i\mu} + \left(\frac{x_3}{x_4}\right)^{i\mu}\right]\, h^2(x_4)\, \left[h^*(x_3)\right]^2.
\eqa
It also follows from the region of integration over $dq$ in (\ref{Leadingoneloop}), in which $q\gg p$. Note that
to calculate the expression for $\kappa_p$ we have used the identity $A_+^* \, A_-^* = A_+ \, A_-$. Thus, for the Euclidian vacuum the leading IR divergence is proportional to $\log\left(1/\eta\eta_0\right) = (t-t_0)$ and is present both in $n_p$ and $\kappa_p$.

For the in-harmonics $h_-(x)$ is proportional to the Bessel function $Y_{i\mu}(x)$. Hence, $h_-(x) \approx A\, x^{-i\mu}$, as $x\to 0$ (past infinity). At the same time, for this case $h_+(x) \approx C_+ \, x^{i\mu} + C_-\, x^{-i\mu}$, as $x\to 0$ (future infinity). Here $A$ and $C_\pm$ are some complex constants related to $A_+$ and $A_-$ via the corresponding Bogolubov rotation.

Then the leading IR contribution in this case looks as:

\bqa
n_p(t) \approx \frac{\l^2 S_{D-2}}{(2\pi)^{D-1}}\, \log\left(\frac{1}{p \, \eta}\right)\, \iint^0_\infty dx_3 \, dx_4 \, \left(x_3 \, x_4\right)^{\frac{D-3}{2}}\times \nn \\ \times \left[\left|C_+\right|^2 \, \left(\frac{x_4}{x_3}\right)^{i\mu} + \left|C_-\right|^2 \, \left(\frac{x_3}{x_4}\right)^{i\mu}\right]\, V_+(x_3)\, V^*_+(x_4) + \nn \\ + \frac{\l^2 S_{D-2}}{(2\pi)^{D-1}}\, \log\left(\frac{1}{p \, \eta_0}\right)\, \iint_0^\infty dx_3 \, dx_4 \, \left(x_3 \, x_4\right)^{\frac{D-3}{2}}\, \left|A\right|^2 \, \left(\frac{x_4}{x_3}\right)^{i\mu} \, V^*_-(x_3)\, V_-(x_4), \nn \\
\kappa_p(t) \approx - \frac{2\,\l^2 S_{D-2}}{(2\pi)^{D-1}}\, \log\left(\frac{1}{p \, \eta}\right)\, C^*_+ \, C^*_- \, \int^0_\infty dx_3 \,\int_\infty^{x_3} dx_4 \times \nn \\ \times \left(x_3 \, x_4\right)^{\frac{D-3}{2}}\, \left[\left(\frac{x_4}{x_3}\right)^{i\mu} + \left(\frac{x_3}{x_4}\right)^{i\mu}\right]\, V_+(x_3)\, V_+^*(x_4).
\eqa
where $V_\pm(x) = h_\pm^2 (x) - \frac{const}{|x|} - \dots$ \cite{Akhmedov:2011pj}. Note that for the in-harmonics $n_p(t)$ depends separately on $t$ and $t_0$ rather than on their difference $(t-t_0)$ and $\kappa_p(t)$ does not diverge as $t_0\to-\infty$. Similar situation also appears for the out-harmonics with one crucial difference that now $\kappa_p$ does diverge as $t_0\to -\infty$, but is finite as $t\to+\infty$. One encounters analogous situation for the one-loop contribution in the scalar QED on the electric pulse background \cite{AkhmedovGuz}.

For the other $\alpha$--harmonics the leading IR contributions to $n_p$ and $\kappa_p$ also, unlike the case of the Euclidian harmonics, depend on $t$ and $t_0$ separately and they both diverge when $t\to +\infty$ and/or when $t_0 \to -\infty$.

\subsection{Kinetic equation in global de Sitter}

Having in mind the discussion of CPP in the previous section, it is not hard to derive the system of kinetic equations in global dS:

\bqa\label{global1}
\frac{dn_j}{dt} = [N\to PN] + 2\,\lambda^2 \sum_{j_1,\vec{m}_1;j_2,\vec{m}_2} {\rm Re} Tr \Biggl\{J_{j,\vec{m};j_1,\vec{m}_1;j_2\vec{m}_2} \, \int_{-\infty}^{+\infty} dt' \, C^*_{j,j_1,j_2}(t) \times \nn \\ \times \left[\left(1+N^*_j\right)\, N_{j_1} \, N_{j_2}^{\phantom{\frac12}} - N^*_j\, \left(1 + N_{j_1}\right)\, \left(1 + N_{j_2}\right)\right](t,t')\Biggr. + \nn \\
+ 2\, J_{j_1,\vec{m}_1;j_2,\vec{m}_2;j\vec{m}} \, \int_{-\infty}^{+\infty} dt'\, C^*_{j_1,j_2,j}(t)\, \left[N^*_{j_1}\, \left(1+N_{j_2}\right) \, \left(1+N_{j}\right)^{\phantom{\frac12}} - \left(1+N^*_{j_1}\right)\, N_{j_2}\, N_{j}\right](t,t') + \nn \\
\Biggl. I_{j,\vec{m};j_1,\vec{m}_1;j_2\vec{m}_2} \, \int_{-\infty}^{+\infty} dt' \, D^*_{j,j_1,j_2}(t)\, \left[\left(1+N_j\right)\, \left(1+N_{j_1}\right) \, \left(1+N_{j_2}\right)^{\phantom{\frac12}} - N_j\, N_{j_1} \, N_{j_2}\right](t,t')\Biggr\}
\eqa
and

\bqa\label{global2}
\frac{d\kappa_j}{dt} = - [N\to PN] - 2\,\lambda^2 \sum_{j_1,\vec{m}_1;j_2,\vec{m}_2} (\vec{m} \to - \vec{m}) Tr \Biggl\{J_{j,\vec{m};j_1,\vec{m}_1;j_2\vec{m}_2} \, \int_{-\infty}^{t} dt' \, C^*_{j,j_1,j_2}(t)\times \nn \\ \times \left[\left(1+N_j\right)\, \left(1+N_{j_1}\right) \, \left(1+ N_{j_2}\right)^{\phantom{\frac12}} - N_j\, N_{j_1}\, N_{j_2}\right](t,t')\Biggr. + \nn \\
+ 2\, J_{j_1,\vec{m}_1;j_2,\vec{m}_2;j\vec{m}} \, \int_{-\infty}^{t} dt' \
C^*_{j_1,j_2,j}(t) \, \left[N^*_{j_1}\, \left(1+N_{j_2}\right) \, \left(1+N_{j}\right)^{\phantom{\frac12}} - \left(1+N^*_{j_1}\right)\, N_{j_2}\, N_{j}\right](t,t') + \nn \\ +
\Biggl. I_{j,\vec{m};j_1,\vec{m}_1;j_2\vec{m}_2} \, \int_{-\infty}^{t} dt' \, D^*_{j,j_1,j_2}(t) \, \left[\left(1+N_j\right)^{\phantom{\frac12}} N^*_{j_1} \, N^*_{j_2} - N_j\, \left(1+N^*_{j_1}\right) \, \left(1+N^*_{j_2}\right)\right](t,t')\Biggr\}
\eqa
The terms $[N\to PN]$ and $(\vec{m} \to - \vec{m})$ have been defined in the subsection B of the section II.
In (\ref{global1}) and (\ref{global2}) instead of the usual flat space $\delta$-functions, which ensure energy-momentum conservation, we have

\bqa
J_{j_1\vec{m}_1;j_2\vec{m}_2;j_3\vec{m}_3} = \int d\Omega Y^*_{j_1\vec{m}_1}\left(\Omega\right)\, Y_{j_2\vec{m}_2}\left(\Omega\right)\, Y_{j_3\vec{m}_3}\left(\Omega\right), \quad C_{j_1,j_2,j_3} = \cosh^{D-1}\left(t\right)\, g^*_{j_1}\, g_{j_2}\, g_{j_3}, \nonumber \\
I_{j_1\vec{m}_1;j_2\vec{m}_2;j_3\vec{m}_3} = \int d\Omega Y_{j_1\vec{m}_1}\left(\Omega\right)\, Y_{j_2\vec{m}_2}\left(\Omega\right)\, Y_{j_3\vec{m}_3}\left(\Omega\right), \quad D_{j_1,j_2,j_3} = \cosh^{D-1}\left(t\right)\, g_{j_1}\, g_{j_2}\, g_{j_3}.
\eqa
Unlike the case of EPP and CPP, we do not yet understand what is the appropriate choice of the harmonics for this system of kinetic equations. Hence, we are not yet in a position to solve it analytically.

\section{Conclusions}

We have derived kinetic equations in CPP, EPP and global dS for the massive scalar fields from the principal series. Our goal is to see whether the kinetic equation in global dS has any stationary solutions.
In EPP and CPP we have found stationary solutions. The possibility of their existence strongly depends on the ratio between the particle production, $\Gamma_2$, and particle decay, $\Gamma_1$, probabilities. This ratio defines the equilibrium particle density and is proportional to the Gibbons-Hawking rate, $e^{-2\pi\mu}$, $\mu = \sqrt{m^2 - \left(\frac{D-1}{2}\right)^2}$. It is suppressed when the particle mass is very big. However, as the mass of the particle $m$ approaches the $(D-1)/2$ bound, the solution is not valid anymore. Furthermore, when such a ratio is grater than one, then the solution in question is unstable under the linearized perturbations \cite{Akhmedov:2012pa}. Thus e.g. for the scalar fields from the complementary series the picture of the particle kinetics can be quite different from what we observe for the principal series.

Furthermore, in global dS particle production and decay rates are different from those in CPP and EPP. As the result the stationary solutions of EPP and CPP found above are not applicable in global dS. The question is if (\ref{global1}) together with (\ref{global2}) have any other stationary solutions.
In this respect it is worth recalling that solutions of these equations for the different choices of the harmonics are related to each other via Bogolyubov transformations. Hence, if the stationary solution does exist, then it means that there is a choice of $g_j(t)$ harmonics for which the system of equations under consideration has a solution with $\kappa_p = 0$. However, we do not see such a choice so far.

The last, but not least, subject which we would like to mention is the UV/IR mixing. In \cite{Krotov:2010ma} it was noticed that UV and IR renormalizations are mixed in the coordinate space correlation functions in global dS. In CPP the situation is similar and, as we discussed above, is quite different from EPP (see also \cite{Kahya:2009sz} for the related discussion in EPP). In fact, the sufficient condition for this mixing is the presence of IR divergences of the type discussed above. However, the necessary condition for such a mixing is the presence of the non--vanishing collision integral. E.g. one can see that in CPP for the case of the stationary solution the mixing disappears and the theory becomes renormalizable.

We conclude that, in the circumstances when the stationary situation can not be reached over the fixed dS background, the resolution of the problem comes only when one takes into account backreaction on the dS metric. Because then the stationary stage should be achieved in the deformed space.

I would like to acknowledge discussions with A.Sadofyev, M.Johnson, I.Morrison and with A.Polyakov. I would like to thank MPI, AEI, Golm, Germany for the hospitality during the final stage of the work on this project. The work was partially supported by the grant "Leading Scientific Schools" No. NSh-6260.2010.2, RFBR-11-02-01227-a and the grant from the Ministry of Education and Science of the Russian Federation, contract No. 14.740.11.0081.

\end{document}